\documentclass{jetpl}
\input epsf.tex
\twocolumn

\lat

\DeclareMathOperator{\cn}{\mathop{\rm cn}\nolimits}
\DeclareMathOperator{\dn}{\mathop{\rm dn}\nolimits}
\DeclareMathOperator{\sn}{\mathop{\rm sn}\nolimits}
\def\be{\begin{equation}}
\def\ee{\end{equation}}
\def\vac#1{{\bf #1}}
\def\bsigma{\mbox{\boldmath $\sigma$}}
\def\br{\noalign{\vskip2pt\hrule height1pt\vskip2pt}}

\title{Numerical investigation of logarithmic corrections in two-dimensional
	spin models}

\rtitle{Numerical investigation of logarithmic corrections\ldots}

\sodtitle{Numerical investigation of logarithmic corrections in two-dimensional
	spin models}

\author{B. \,Berche$^{(*)}$,  
	L.\,N. \,Shchur$^{(*,**)}$
	}

\rauthor{B. \,Berche, L.\,N. \,Shchur}

\sodauthor{Berche, Shchur}

\address{$^{(*)}$ Laboratoire de Physique des Mat\'eriaux, UMR CNRS 7556,\\ 
  	Universit\'e Henri Poincar\'e, Nancy
  	1, B.P. 239,
  	F-54506  Vand\oe uvre les Nancy Cedex, France\\~\\
  	$^{(**)}$ Landau Institute for Theoretical Physics, \\
	Russian Academy of Sciences,
	Chernogolovka 142432, Russia}

\dates{January 19th, 2004}{*}

\abstract{
	The analysis of correlation function data obtained by
	Monte Carlo simulations of the two-dimensional 4-state Potts model,
	XY model, and self-dual disordered Ising model 
	at criticality are presented.
	We study the logarithmic corrections to the algebraic decay 
	exhibited in these models. A conformal mapping is used to relate the 
	finite-geometry information to that of the infinite plane.
	Extraction of the leading singularity is altered by the expected 
	logarithmic corrections, and we show numerically that both leading
	and correction terms are mutually consistent.
}


\begin{document}

\maketitle


A second order phase transition occurs at very special points in the 
parameter space of a model, i.e. at a fixed point of the 
renormalization equations.
The eigenvalues of the linearized renormalization equations define the
scaling dimensions of the corresponding directions related to the 
scaling fields. Positive eigenvalues are associated to relevant scaling
fields while negative ones correspond to irrelevant fields.  
In some cases, there may exist a line of fixed points along which 
critical exponents are varying. This occurs when a marginal field (with
vanishing scaling dimension) is identified in the model.
The scaling dimensions completely characterize the critical properties
of the model, e.g. the power laws of the physical quantities. 
As an example, the correlation functions exhibit an algebraic decay at 
the fixed point, e.g. for a scaling field density $\phi(\vac r)$,
\be	G_\phi(\vac r_1,\vac r_2)=
	\langle\phi(\vac r_1)\phi(\vac r_2)\rangle
	\sim |\vac r_1-\vac r_2|^{-\eta_\phi}
	\label{eq1} 
\ee
in two dimensions ($\eta_\phi$ should be replaced by $d-2+\eta_\phi$ in
arbitrary dimension).
In some special cases, this simple behaviour is modified by multiplicative
logarithmic terms,
\be	G_\phi(\vac r_1,\vac r_2)
	\sim |\vac r_1-\vac r_2|^{-\eta_\phi}\times
	\ln^{\theta_\phi} |\vac r_1-\vac r_2|.
	\label{eq2} 
\ee
This situation occurs at the end of a line of fixed points or when the system
is perturbed by a marginally irrelevant operator.
Examples in two dimensions
are given by 4-state Potts model,  XY model, or disordered Ising model.

In the case of the $q$-state 
Potts model (from now on, we consider the case of the
two dimensional problem only), discrete spin variables
$\sigma_w=1,2\dots q$ are located at the sites $w$ of a square lattice
($\mu$ specifies the unit lattice vector in the two directions) and interact
with nearest neighbours,
\be	-\frac {H_q}{k_BT}=K\sum_w\sum_\mu\delta_{\sigma_w,\sigma_{w+\mu}}.
	\label{eq3}
\ee
The value of the number of states per spin, $q$, can be seen as a parameter
of the model. The special case $q\to 1$ corresponds to the percolation problem
and exhibits a second order phase transition. The value $q=2$ 
corresponds to the Ising model, so  when 
$q$ varies the universality class changes. The transition is of second order
as long as $q\le 4$ and the value $q=4$ coincides with the end of the
line of fixed points. Along this line, 
the exponent of the {\em order parameter
correlation function} varies according 
to~\cite{denNijs79ETAL}
\be \eta_\sigma=\frac{(m+3)(m-1)}{4m(m+1)},\label{eq4}\ee
where the value of 
$m=\pi/{\rm cos}^{-1}({\small\frac 12}\sqrt q)-1$ parametrizes the number of
states.  At $q=4$, the correlation function exponent takes the
value $\eta_\sigma(q=4)={\small\frac 14}$.
Above $q=4$ the transition becomes of first order.
The transition temperature of the $2d$ Potts model
is exactly known from duality requirements~\cite{Wu82},
$K_c=\ln(1+\sqrt q)$.

The two-dimensional classical XY model is another example of such a
scenario, although the transition is in its very nature quite different.
The Hamiltonian now describes the interaction of classical
two dimensional unit spins, $\bsigma_w=(\cos\theta_w,\sin\theta_w)$,
\be	-\frac {H_{{\rm XY}}}{k_BT}=K\sum_w\sum_\mu \bsigma_w\cdot\bsigma_{w+\mu}.
	\label{eq5}
\ee
At low temperature, as a consequence of the Mermin-Wagner-Hohenberg 
theorem~\cite{MerminWagner66ETAL}, 
there is no finite macroscopic magnetization, but the
spin-spin correlation function decays algebraically in the so-called 
quasi-long-range-ordered phase. The temperature plays the role of a marginal 
field and the decay exponent $\eta_\sigma$ continuously increases up to
a limiting value $\eta_\sigma({\rm XY})={\small\frac 14}$ 
at a transition temperature named after 
Berezinskii-Kosterlitz-Thouless 
(BKT)~\cite{Berezinskii71ETAL}. 
In this low temperature critical phase,
ordering is prevented by collective excitations, spin waves, and localized 
excitations, vortices, which appear in increasing number as the temperature
is increased. These latter topological defects are bounded in pairs and the
transition to a completly disordered phase with exponential decay of the
correlations is reached at the BKT temperature when unbinding of the pairs
occurs~\cite{Villain75ETAL}.  
The BKT transition thus corresponds to the end of a continuous
line of fixed points in the low temperature phase.
The transition temperature $K_c\simeq 0.893$, as well as the  
expression of the temperature dependence of the exponent $\eta_\sigma$
are not known exactly. 
The mechanism of the transition may also be understood
from the role of the vortex chemical potential. It is a relevant variable 
in the low temperature phase, which becomes marginally irrelevant at
the BKT transition, thus producing the essential singularities and
the logarithmic corrections.

A third example is provided by the random bond Ising model in two dimensions.
According to Harris criterion~\cite{Harris74}, 
quenched disorder is a relevant variable
when the specific heat exponent of the pure system under consideration is 
positive. In this case, a new fixed point leading to a new universality class 
is expected. In the case of the $2d$ Ising model, since $\alpha=0$ in the
pure model, randomness is only a marginal variable which could either
produce a continuous variation of the exponents with the amplitude of disorder
or logarithmic corrections to the unchanged leading critical behaviour if
the disorder is eventually marginally irrelevant.
After an interesting debate in the 
eighties~\cite{DotsenkoDotsenko83}, the second 
scenario has been recognized
to be 
correct~\cite{Shalaev84ETAL} 
and the 
correlation function exponent keeps its value 
$\eta_\sigma({\rm RBIM})={\small\frac 14}$. 
The Hamiltonian is the one given in equation~(\ref{eq3}),
with $q=2$ and with random nearest neighbour interactions $K_{w,\mu}$
along the bonds between sites $w$ and $w+\mu$.
When these couplings are taken from a binary probability distribution
with equal probabilities for both strengths,
${\cal P}[K_{w,\mu}]={\small\frac 12}\prod_{w,\mu}[\delta(K_{w,\mu}-K_1)
+\delta(K_{w,\mu}-K_2)]$, the critical temperature follows from duality,
$(\exp (K_1^c)-1)(\exp (K_2^c)-1)=q$ ($=2$ here).

All these models have the common property that the order parameter correlation
function $G_\sigma$ is expected, 
from renormalization group arguments, to behave
according to 
equation~(\ref{eq2}) with $\theta_\sigma={\small\frac 18}$ (XY model) and
$-{\small\frac 18}$ (Potts model and 
RBIM)~\cite{Shalaev84ETAL,CardyNauenbergScalapino80,AmitGoldschmidtGrinstein80}.
One may then define a local effective exponent $\eta_{{\rm eff}}(r)$ 
($r=|\vac r_1-\vac r_2|$),
\be 
	\eta_{{\rm eff}}(r)=-\frac{{\rm d}\ln G_\sigma(r)}{{\rm d}\ln r}
	=\eta_\sigma-\frac{\theta_\sigma}{\ln r}\label{eq6}.
\ee
To make this formula more explicit, let us put some typical numbers.
Suppose that we want to produce numerical simulations, in order to check
equation~(\ref{eq2}), with let say a relative accuracy of 
$\Delta\eta/\eta_\sigma=\frac{|\theta_\sigma|}{\eta_\sigma\ln r}=10^{-2}$.
The exponent $|\theta_\sigma|$  
might be estimated to be of the order of the value of 
$\eta_\sigma$, so that one has to reach values of $r$ as large as
$r\simeq\exp((\Delta\eta/\eta_\sigma)^{-1})\simeq 10^{43}$!
Even with a value of $|\theta_\sigma|={\small \frac 18}$, 
one still needs a sample
much larger than $L>10^{21}$ in order to reach distances between spins
which are large enough to become sensitive to the presence of the log in
equation~(\ref{eq2}). 
This is definitely not possible and that might be the reason why it was
almost impossible to produce 
reliable data\,\footnote{Both by simulations and by
series expansions.} 
to corroborate
equation~(\ref{eq2}). The strategy is thus to accept the leading behaviour
(the value of $\eta_\sigma$ as predicted by RG), 
and to extract the value of $\theta_\sigma$.
This is also a difficult route, and contradictory results
were reported in the literature.
In the case of the XY model for example, fixing 
$\eta_\sigma={\small\frac 14}$ and using Monte Carlo simulations 
(susceptibility and correlation length data), 
Kenna and 
Irving~\cite{KennaIrving95ETAL} 
reported a 
value close to
$\theta_\sigma\simeq 0.046(20)$, Janke~\cite{Janke97} obtained 
$\theta_\sigma\simeq 0.054(2)$ from data at criticality, or
$\theta_\sigma\simeq -0.112(4)$ using high temperature data.
Also with high temperature data, 
Patrascioiu and Seiler~\cite{PatrascioiuSeiler96} obtained
$\theta_\sigma\simeq -0.154(92)$ and  from strong coupling expansion
Campostrini et al.~\cite{CampostriniPelissettoRossiVicari96} 
had results depending on the lattice symmetry,
$\theta_\sigma\simeq -0.090(6)$
to $-0.084(12)$\,\footnote{In the literature, $\theta_\sigma$ is often 
referred to as $-2r$.}.
There are even more controversial results reported by 
Balog et 
al.~\cite{Balog01ETAL} 
who suggest
$\theta_\sigma=0$ plus additive corrections, or 
Kim~\cite{Kim96} who is in favor of
ordinary scaling rather than essential singularities.
We are not aware of any direct verification of the presence of logarithmic
terms directly in the correlation function in the case of 4-state Potts
model or disordered Ising model, but logarithmic terms have been found
to be compatible with finite-size scaling data of susceptibility and 
specific heat~\cite{SalasSokal97}
in the 4-state Potts model and  
in the random bond Ising model~\cite{WangSelkeDotsenkoAndreichenko90}. 

There are conceptual difficulties with the above-mentioned approach applied
to numerical simulations. First, the maximum available linear extent of a 
lattice is of the order of $L=10^3$, and due to boundary effects, only a 
fraction (let say $1/4$th of the lattice) can be used. As a 
consequence, the value of $\eta_\sigma$ is strongly altered by the
log-correction in eq.~(\ref{eq6})\,\footnote{With 
$\theta_\sigma=\pm{\small\frac 18}$ and
simulations available up to relative distances as large as $r=500$, one can at
most reach values of $\eta_\sigma$ in the range $0.23-0.27$.}.
Keeping a fixed $\eta_\sigma={\small\frac 14}$ means that we attribute
all the deviation of numerical data to the existence of $\theta_\sigma$, but
again, this correction is only an effective one, since the log term in the
correlation function~(\ref{eq2}) is only the first of a series and the 
next term (see e.g. Ref.~\cite{AmitGoldschmidtGrinstein80,SalasSokal97}) 
will affect in a similar manner
the value of $\theta_\sigma$ and make it an effective 
one\,\footnote{Using the expression $G(r)\times r^{1/4}\sim
\ln^{1/8}r\times(1+\frac{1}{16}\frac{\ln\ln r}{\ln r})$ given in 
Ref.~\cite{AmitGoldschmidtGrinstein80} for the BKT transition and 
again with $r\simeq 500$,
we get $\theta_\sigma(r)=\theta_\sigma+\frac{1}{16}
\frac{1-\ln\ln r}{\ln r+(\ln\ln r)/16}
\simeq 0.117$.}. 
A second objection comes from the appearence of a typical length scale
which has strong influence in the correction to scale-invariant behaviour.
Usually, one does not 
have to take care to the unit length, since the physical quantities
exhibit scale invariance at criticality. Strictly speaking, 
this is no longer true when a correction to scaling (e.g. a logarithm) 
is present, and the above expressions
for the correlation functions should be rewritten with respect to some 
scale factor $a$. This factor is
related to the lattice spacing, but as well to some physics 
of the problem, like the size of vortex pairs in the XY model or the typical
disorder length in the RBIM, it is thus
non universal and depends on the model.
The local exponent becomes
\be 
	\eta_{{\rm eff}}(r)
	=\eta_\sigma-\frac{\theta_\sigma}{\ln r/a}\label{eq8}.
\ee
What is important to notice here is the fact that the value of $a$
determines the amplitude of variations of the effective exponent 
$\eta_{{\rm eff}}(r)$.
According to these observations, it is probably already
a good result to only predict a correct sign for the correction 
exponent $\theta_\sigma$, and we have no stronger ambition in this paper.

We propose  a different approach which is almost free from
boundary effects and enable us to work with the asymptotic expression of
the correlation function in the infinite plane. 
The simulations\,\footnote{Standard Wolff cluster algorithms.} 
are performed inside a finite system,
but the functional expression of the
correlation function inside such a system is 
predicted by a convenient conformal mapping. 
This method has been applied with success to magnetization profiles
in the case of the pure XY
model~\cite{ResStraley00ETAL} 
and was extensively used in the case of 
disordered Potts
models~\cite{BercheChatelain02} 
in two dimensions where 
it was shown to provide quite accurate
results.
More problematic is the fact that we apply a 
method which is known to be valid at a really scale-invariant fixed point,
i.e. in the absence of corrections to scaling which break (at the correction 
level) dilatation symmetry.
In the following, we consider
systems of reasonable sizes ($L$ up to 256), the asymptotic regime
$r\to\infty$, $G_\sigma(r)\sim r^{-\eta_\sigma}$, 
is thus far from being reached and the variable $\ln r$ only 
varies within a narrow range. Accordingly, eq.~(\ref{eq2}) might be
replaced by an algebraic decay with an effective exponent,
$G_\sigma(r)\sim r^{-\eta_{{\rm eff}}}\label{eqGeff},$ 
and displays scale invariance, at least in the range of distances
under consideration.

\begin{figure} [h]
\vspace{0.2cm}
        \epsfysize=3.6cm
        \mbox{\epsfbox{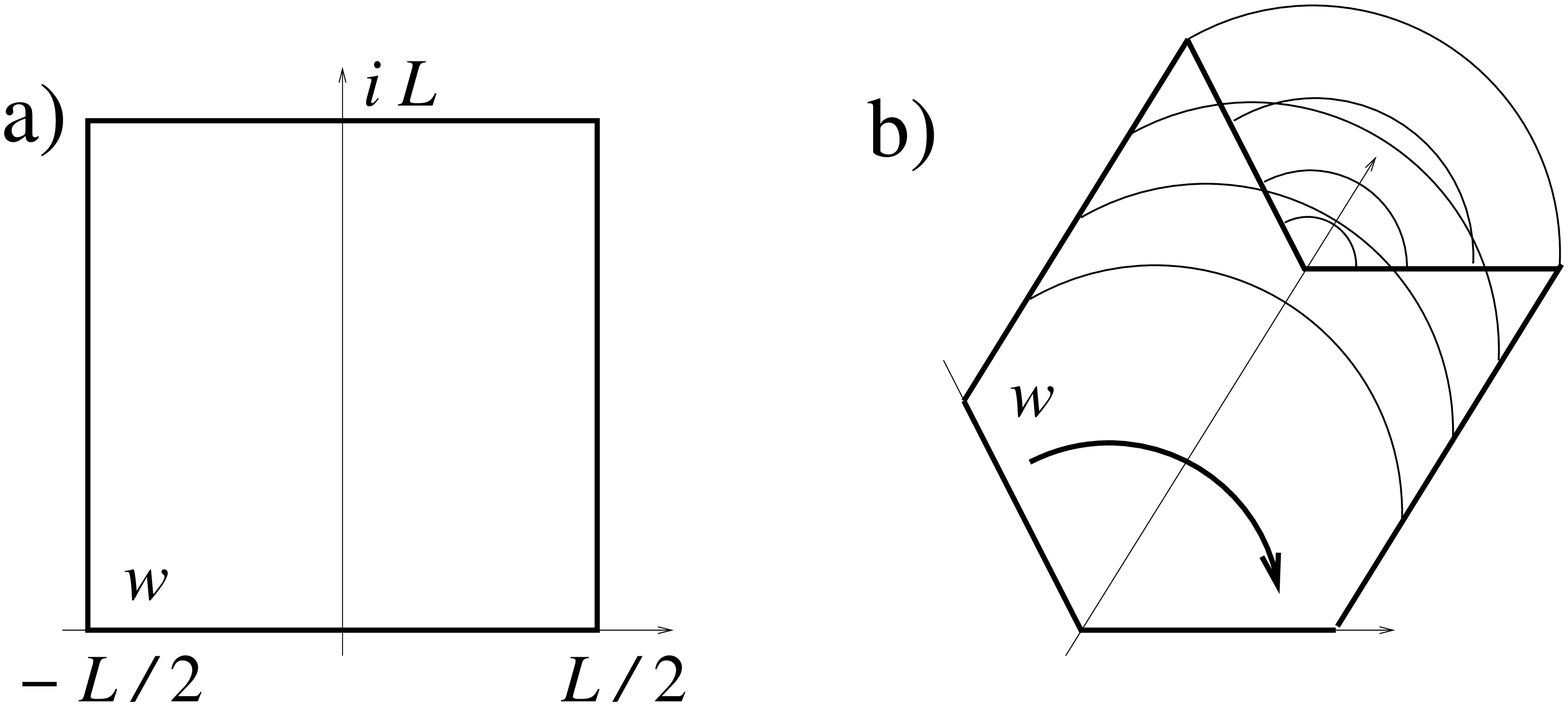}\hfil\ }
	\epsfysize=4.0cm
	\begin{center}\mbox{\epsfbox{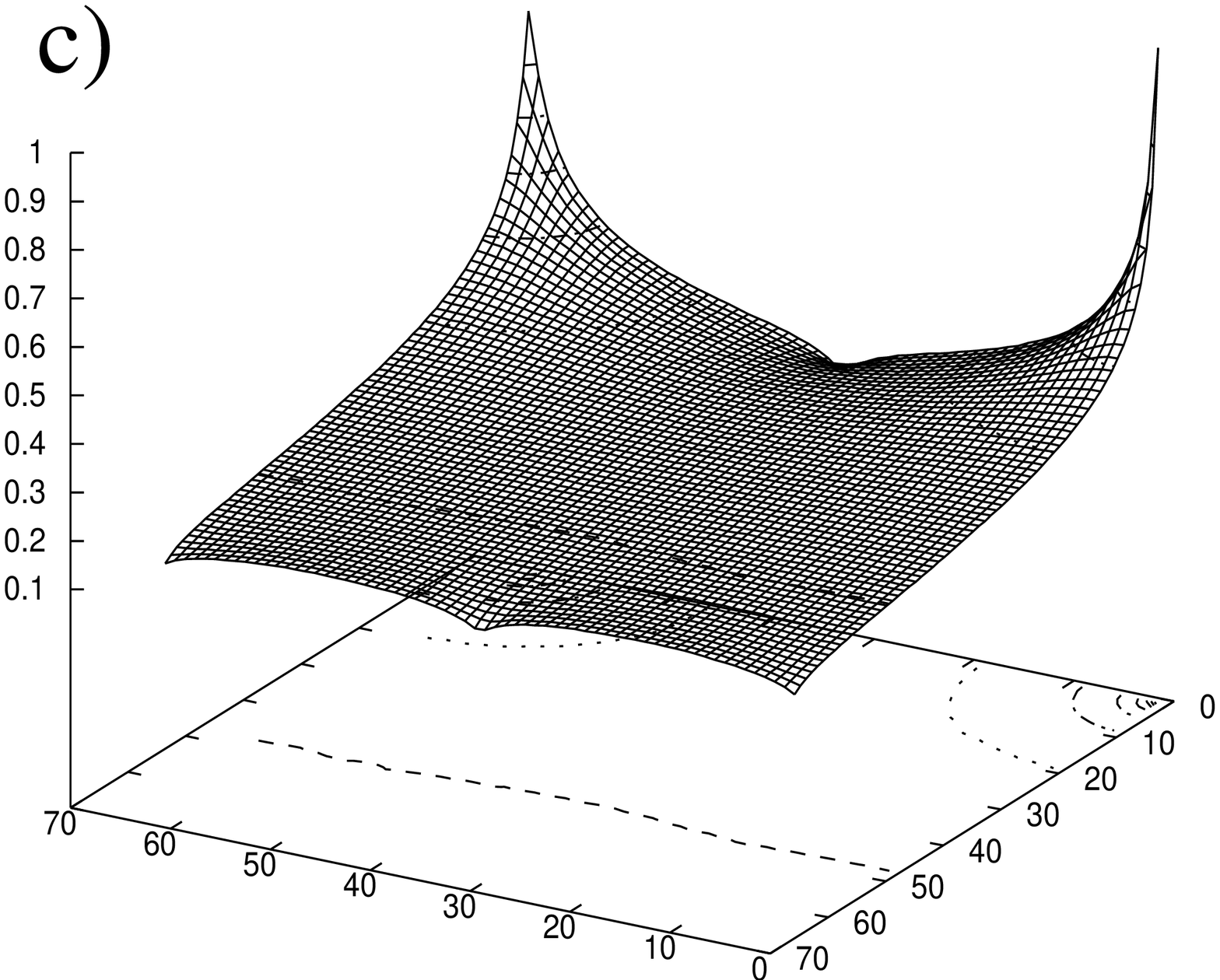}}
        \end{center}
        \caption{Fig. 1: a) Conformal mapping of the infinite 
	complex plane inside
	a square $w$. b) Sketch which shows how the boundary
	conditions (BC) follow from the folding (``pillow'' geometry).
	c) Example of the profile of the correlation
	function between the upper right corner and other points in the
	square with these particular BC.}
        \label{fig1}  \vskip -0cm
\end{figure}

\noindent - The upper complex plane $\Im{\rm m}\ \! \zeta\ge 0$ is 
mapped inside a square 
$-L/2\le \Re{\rm e}\ \! w\le L$, $0\le \Im{\rm m}\ \! w\le L$  
through 
the Schwarz-Christoffel 
conformal mapping $\zeta=\sn\frac{2{\rm K}w}{L}$. 

\noindent - In order to relate this
finite geometry to an original infinite plane (with complex variable $z$),
one may use the Schwarz transformation $\zeta=z^{1/2}$ which has the effect 
of a folding of opposite edges of the 
square two by two~\cite{BurkhardtDerrida85} as 
shown in figure~\ref{fig1}.

There, the effect
of the conformal mapping,
\be 
G_\sigma(w_1,w_2)\sim |w'(z_1)|^{-\frac 12\eta_\sigma}
|w'(z_2)|^{-\frac 12\eta_\sigma}
G_\sigma(z_1,z_2), 
\ee
is just to define a rescaled
distance variable, called $\kappa(w_1,w_2)$, in terms
of which one recovers inside the square with these special boundary 
conditions, a
simple power law for the correlation function:
\begin{eqnarray}
G_\sigma(w_1,w_2)
&\sim&[\kappa(w_1,w_2)]^{-\eta_{{\rm eff}}},\\
\kappa(w_1,w_2)&=&|w'(z_1)|^{\frac 12}
|w'(z_2)|^{\frac 12}|z_1-z_2|,
\label{eq7}
\end{eqnarray}
where $|w'(z)|= \frac{L}{4{\rm K}}\left| 
 \cn (2{\rm K}w/L)
 \dn (2{\rm K}w/L)
 \sn (2{\rm K}w/L)\right|^{-1}$
and  
$G_\sigma(z_1,z_2)$
is the correlation function
in the plane (with $z=\sn^2(2{\rm K}w/L)$). 
Here, $\cn x$, $\dn x$ and $\sn x$ are the Jacobi elliptic functions, 
$L$ the linear size of the lattice,
 and ${\rm K}\simeq 1.58255$ 
is a constant related to the aspect ratio 
of the system.

\begin{figure} [h]
\vspace{0.2cm}
        \epsfxsize=8cm
        \begin{center}
        \mbox{\epsfbox{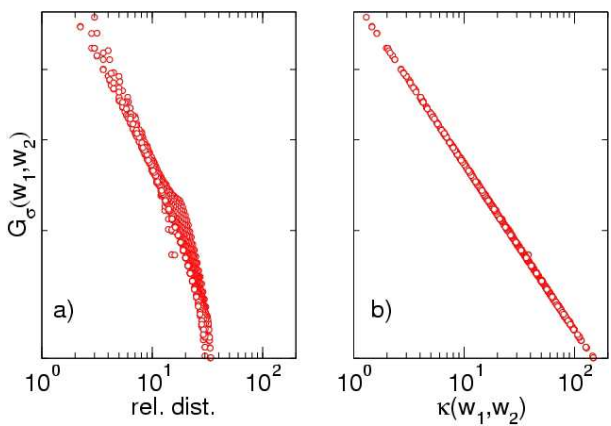}\hfil\ }
	\end{center}
        \caption{Fig. 2: a) Log-log 
	plot of the correlation function $G_\sigma(w_1,w_2)$ of the XY model
	{\it vs} the relative distance in the original 
	square geometry for a system of size
	$32^2$. 
	b) Same data plotted now {\it vs} the rescaled
	relative distance $\kappa(w_1,w_2)$.
	}
        \label{fig_w-et-z}  \vskip -0cm
\end{figure}

\begin{figure} [h]
\vspace{0.2cm}
        \epsfxsize=8cm
        \begin{center}
        \mbox{\epsfbox{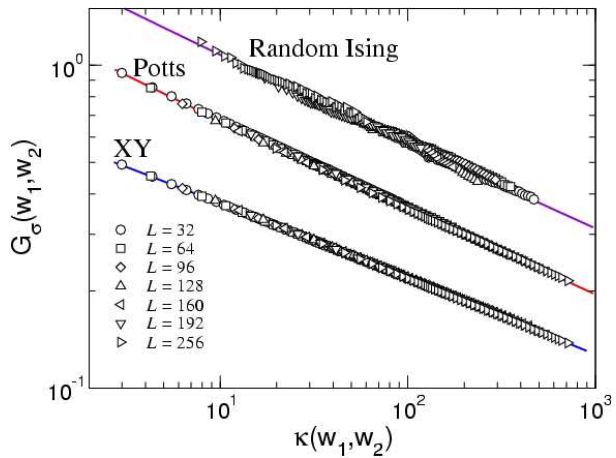}\hfil\ }
	\end{center}
        \caption{Fig. 3: Log-log plot of the correlation function 
	inside the square, 
	$G_\sigma(w_1,w_2)$
	as a function of the convenient rescaled variable 
	$\kappa(w_1,w_2)$ (for sizes from
	$L=32$ to $L=256$ for the three models under consideration). }
        \label{fig2}  \vskip -0cm
\end{figure}

The main advantage of this technique is that one lattice size 
$L$ is in principle 
sufficient (provided it is large enough), since the shape effects are 
included in the conformal mapping and
the method is not much sensitive to finite-size effects. The effect of
discretization of the lattice is only appearent at the scale of a few 
lattice spacings. One more advantage is the fact that all the information
encoded in the correlation function is used, since {\em all} 
the points $w$ inside the square enter the fit (see fig.~\ref{fig_w-et-z}). 
Now, as we noticed above, it is necessary to take into account the existence
of the logarithmic term if we want to understand the leading singularity.
In order to emphasize this comment,
let us discuss briefly the results presented in figure~\ref{fig2} for
$L=32$ to $256$ in the case of XY, 4-state Potts and random-bond Ising
models.
We show the log-log plot of 
$G_\sigma(w_1,w_2)$ with respect to the rescaled distance 
$\kappa(w_1,w_2)$  
for the three models under
consideration. One observes the remarkable linear regime (on this log-log 
scale) over the whole range of variables, and in particular no boundary
effects, as these were included in the conformal mapping. 
Nevertheless, while the expected
slopes should all be equal to the same $\eta_\sigma=\frac 14$, 
a deviation from this value is suspected.
Then, a power
law fit leads to  leading singularities with exponents\,\footnote{Statistics 
over the 7 values of size $L$ for each model.}
$\eta_{{\rm eff}}({\rm XY})\simeq 0.233(3)$,  
$\eta_{{\rm eff}}({\rm PM})\simeq 0.264(6)$ and 
$\eta_{{\rm eff}}({\rm RBIM})\simeq 0.265(14)$ 
(there is a slight variation, depending on the size). All these results are
in poor agreement with the exact result $\frac 14$. This is a clear 
evidence that the logarithmic correction has to be taken into account.

\begin{figure} [h]
\vspace{0.2cm}
        \epsfxsize=8cm
        \begin{center}
        \mbox{\epsfbox{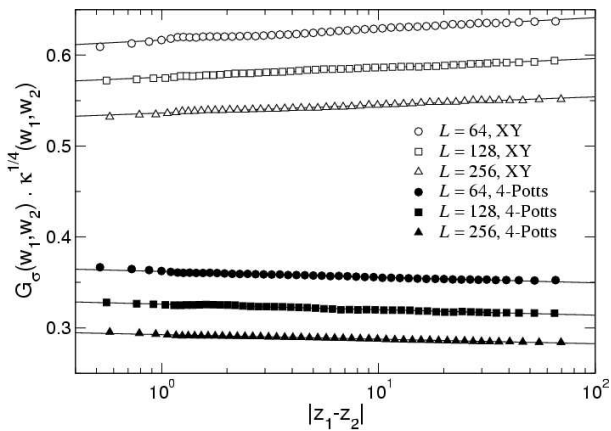}\hfil\ }
	\end{center}
        \caption{Fig. 4: Semi-log plot of the rescaled correlation function
	for XY and Potts models {\it vs} relative distance in the plane 
	geometry ($L=256$). The slope $B_1$ is positive for XY model and 
	negative in the 4-state Potts model.}
        \label{fig3}  \vskip -0cm
\end{figure}

Figure~\ref{fig3} shows 
a semi-log plot of the {\em rescaled} correlation function
$f(|z_{12}|)=G_\sigma(w_1,w_2)\times[\kappa(w_1,w_2)]^{\eta_\sigma}$ 
against
the relative distance $|z_2-z_1|$ in the infinite plane geometry. One
observes empirically that the expected
behaviour
$f(|z_{12}|)\sim A(\ln(|z_{12}|/a))^{\theta_\sigma}$ according to 
eq.~(\ref{eq2})
is in fact linear in this
scale, that is 
\be f(|z_{12}|)\sim B_0+B_1\ln |z_{12}|\label{eq10}.\ee 
This is coherent with the logarithmic correction provided that the 
dimensionless inverse typical length scale
$a^{-1}$ is larger than any of the accesible relative dimensionless
distances at the
sizes available, $a^{-1}\gg max|z_{12}|$. Under these conditions, the
logarithmic correction yields
\be 
f(|z_{12}|)\sim A(\ln(a^{-1}))^{\theta_\sigma}\left(1+
\frac{\theta_\sigma}{\ln(a^{-1})}
\ln|z_{12}|\right),
\ee 
{\it i.e.} $\theta_\sigma$ appears in the
ratio  $B_1/B_0=\theta_\sigma/\ln(a^{-1})$. The slope
$B_1$ in eq.~(\ref{eq10}) is positive if 
$\theta_\sigma>0$ and negative otherwise.

\noindent As we have mentioned previously,  the 
accessible lattice sizes are too small, so that
we do not expect any precise determination
of the correction exponent $\theta_\sigma$. 
More dramatic is the fact that the 
$\theta_\sigma$ exponent appears in our expressions mixed with the non 
universal length scale $a$ from which a reliable value would hardly 
be extracted.
Nevertheless, what is shown in 
Fig.~\ref{fig3} is that a logarithmic correction is consistent with the data
for both XY and 4-state Potts models, that the signs of the exponents of
the log terms are opposite for both models and that their absolute values
are quite close to each other (see table~1). 
In the case of the disordered Ising model, 
a quantitative analysis is made difficult due to the fluctuations 
introduced by the disorder average, but the leading exponent $\eta_{\rm eff}$
was compatible with a negative correction exponent $\theta_\sigma$, as
expected.

\begin{table}
\small
\caption{Tab. 1: Values of the effective 
	exponents deduced from fits of the curves
	shown in figs.~\ref{fig2} and \ref{fig3}. In the case of RBIM, 
	the value of $B_1/B_0$ is roughly 3 times larger than for Potts
	(same sign), but strongly fluctuating.
	}
\vglue0mm\begin{center}
\begin{tabular}{@{}*{7}{l}}
\br
&\multicolumn{3}{c}{$\eta_{{\rm eff}}$} && 
\multicolumn{2}{c}{$\theta_\sigma/\ln (a^{-1})$}
\\
\cline{2-4} \cline{6-7}
 $L$ & XY & Potts & RBIM && XY & Potts \\
\hline
$64$ & $0.231$ & $0.268$ & $0.259$ && $0.0087$ & $-0.0075$ \\
\noalign{\vskip-2mm}
$128$ & $0.233$ & $0.266$ & $0.287$ && $0.0077$ & $-0.0081$ \\
\noalign{\vskip-2mm}
$256$ & $0.234$ & $0.260$ & $0.255$ && $0.0072$ &  $-0.0074$ \\
\br
\end{tabular}
\\
\end{center}
\label{tab1}
\end{table}

{\bf Acknowledgment:} We gratefully 
acknowledge Ralph Kenna for stimulating discussions.
We thank the Twinning program between the CNRS and the Landau Institute 
which made possible this pleasant cooperation. Partial support from RFBR is
acknowledged.


\end{document}